\documentclass{iau}

\usepackage{amsmath}
\usepackage{graphicx}
\usepackage{multirow}

\begin{document}

\lefttitle{Luk\'a\v{s} Likav\v{c}an}
\righttitle{Substrate-Agnostic 3x: Biosignatures, Technosignatures, Ecologies}

\jnlPage{1}{7}
\jnlDoiYr{2026}
\doival{10.1017/xxxxx}

\aopheadtitle{Proceedings IAU Symposium No.~404, 2026}
\editors{J. Haqq-Misra \& R. Kopparapu, eds.}

\title{Substrate-Agnostic 3x: Biosignatures, Technosignatures, Ecologies}

\author{Luk\'a\v{s} Likav\v{c}an}
\affiliation{Institute of Philosophy, Slovak Academy of Sciences}

\begin{abstract}
Substrate-agnostic perspectives are currently attracting increased attention. For example, it has become customary to refer to agnostic biosignatures to reflect the range of alternative extraterrestrial biospheres and to account for the deeper philosophical dependence of candidate biosignatures on the underlying theory of life. Analogously, one can formulate a concept of agnostic technosignatures, reflecting that the more we expand the search for technosignatures, the more we invite theories of technology that undo the terrestrial bias. For this reason, this paper argues that there exists a strong theoretical justification for an integrated study of technosignatures and biosignatures, articulated in a unified perspective on substrate-agnostic ecologies. The paper introduces the concept of substrate-agnostic ecology as an abstraction unconstrained by terrestrial circumstances, anchored instead in a functional understanding of agent-ecology coupling provided by niche construction theory.\end{abstract}


\maketitle

\section{Introduction}

This paper outlines an argument for adopting a philosophical standpoint called \textit{substrate-agnostic ecology}. This standpoint reconceptualizes the relationship between biosignatures, technosignatures, and ecological thinking by performing a triple abstraction: one in how we define life, another in how we define technology, and a third in how we think about their convergence within ecologies of niche-constructing agents. It proceeds by outlining the shifting sands of philosophical stakes behind the contemporary search for biosignatures, rearticulates the takeaways of this outline in context of the search for technosignatures, and offers a pathway to unify the treatment of both domains by proposing a new theoretical vocabulary, thus taking one step further the arguments first developed in \cite{likavcan_long_2025} and \cite{likavcan_grass_2025}.

\section{Agnostic biosignatures}

Following David Dun\'er, biosignatures must be analyzed as \textit{signs}, i.e., semiotically. In this respect, they can be treated as indices of present or past extraterrestrial life: spectral, chemical, or morphological traces that we interpret as expressions of living systems. Importantly, the interpretation of biosignatures is always dependent on understanding ``the physical processes that let us link the signifier with the signified'' \citep[p.~50]{duner_semiotics_2018}, which implies that any evidence of life on a distant planet is in principle ambiguous: for every (even plainly hypothetical) biologically produced chemical compound, one is required first to rule out potential abiotic processes that can synthesize the same chemical.

The entire inferential machinery of biosignature science depends on what we take life to be, meaning that in order to be in possession of reliable biosignatures, we need a reliable, general theory of life (one which would enable us to abstract from the specifics of terrestrial environments) \citep{smith_life_2023}. Instead of waiting for such a theory of life to emerge, agnostic biosignatures can compensate for this deficit. One data point that can guide us here is NASA's widely used working definition: ``life is a self-sustaining system capable of Darwinian evolution'' \citep{nasa_nasa_2025}. This definition singles out some potentially generic features of life, such as metabolism (``self-sustaining system'') or evolution (as movement in complexity space). What is important, though, is that by tying the concept of life to its capacity to evolve, this definition indicates a vector of complexity: as life evolves, its progress through the evolutionary space of possibilities is indexed by the increase in the potentially measurable complexity of the planetary environment. If such a generic feature can be captured mathematically, we would have a good candidate for a biosignature liberated from any underlying theory of life.

Indeed, some proposed agnostic biosignatures rely on complexity metrics. As an example, consider the proposal of \cite{bartlett_assessing_2022}, which measures the statistical complexity of time-series data from exoplanet observations using \textit{epsilon machines}. This approach eliminates the need for spatially resolved data, as it simply uses a time series of measurements across numerous wavelengths from an instrument such as an infrared telescope. These inputs are processed by the epsilon machine: an algorithm that generates a compressed model capable of producing a time series statistically equivalent to the input time series (hence the model indexes the compressibility of the input, that is, the minimal algorithm that can reproduce it). 

Another candidate approach is the \textit{assembly index} proposed by \cite{sharma_assembly_2023}: it measures the \textit{temporal depth} of the given system (its depth in \textit{assembly space}), defined as the number of steps needed to construct it. In this logic, the temporal depth of the water molecule is several orders of magnitude lower than that of the Sun, Jupiter, and Earth, respectively, leading \cite{walker_life_2024} to claim that, in terms of its history, Earth is the largest object in the known universe. Since assembly theory is not a theory of life but a theory for quantifying universal features of evolved objects, it is an agnostic biosignature (just as statistical complexity is). However, it also comes with a caveat: assembly theory may be theoretically robust and interesting, but it remains to be seen how it can be operationalized into a remote observation strategy for targeted SETI search. As far as the epsilon machines approach identifies relevant data (the time series of measurements across multiple wavelengths), it currently has an upper hand.

Agnostic biosignatures provide a philosophical basis for what could be labeled the first level of substrate agnosticism. Referring to the proposal by \cite{wong_searching_2022} and \cite{bartlett_defining_2020} to replace the narrow concept of \textit{life} with the more generous concept of \textit{lyfe}, one can treat different complexity measures as a way to characterize a more generic quality of planets than habitability would normally be: namely \textit{genesity}, defined as the creative capacity of a planet to generate novel forms of existence, regardless of their specific biochemical implementation.

\section{Agnostic technosignatures}

Extending the line of reasoning just proposed regarding biosignatures, the second gesture of abstraction should generalize not just upon substrates of life but also upon substrates of intelligence (basically building a theory-independent approach to the detectability of extraterrestrial intelligence). By the same token, it should provide a basis for a substrate-agnostic approach to technology (often assumed to be a detectable manifestation of intelligence) and, hence, expand the conceptualization of technosignatures beyond the narrow confines of terrestrial biases. Additionally, one has to probe the very relationship between technosignatures and biosignatures here, since any technosignature can be treated as a substrate-agnostic biosignature. This amounts to saying that there may be no sufficient reasons to restrict certain genres of technology to certain genres of biochemical substrates, and vice versa. That some CHNOPS-based lifeforms on Earth seem to be heavily invested today in building machines made of iron or silicon is a contingent fact, and the same holds for the structural features of technology. To put it differently, to name something as ``artificial'' requires understanding the context in which the supposedly ``artificial'' intervention takes place.

The effort to abstract from terrestrial particularities resembles some fundamental theoretical considerations in archaeology, design, and architecture. These disciplines also frequently grapple with the definition of artificial structures. Interestingly, designers and architects struggle to define their own scope today because they must confront the nonhuman dimension of artifactual production, whether by animals or machines. This leads to abandoning the parochial emphasis on human examples of purposeful design, currently catalyzed by the rapid onboarding of AI as a creative agent on its own terms. Elsewhere, archaeology finds its epistemological and methodological foundations at stake, having to keep reassessing what counts as reliable evidence of past human cultures \citep[cf.][]{cleland_epistemological_2007}. To some extent, archaeology's position is easier than that of architecture and design, since the discipline explicitly constrains its scope to human cultures. Yet, it must also constantly shrug off potential biases inherent in extrapolations from the known pasts (and presents) to keep enough wiggle room for genuine discoveries.

This line of reasoning brings us to the classic formulation by \cite[p.~20]{tarter_evolution_2007} of technosignatures as evidence of some technology that modifies its environment in detectable ways. The emphasis here falls on \textit{environmental modification}, not on the specific artifacts or engineering feats we might associate with terrestrial technology. Of course, whether we talk about astroengineering or Dyson spheres, these are environmental modifications too (on the scale of cosmic environments), but they are still just extrapolations of past, current, or likely future technologies deployed by the planetary community of humans (no matter how far-fetched). Sticking with them means continuing the tacit universalization of the technical evolution we are familiar with from Earth, thereby ramifying the concept of technology as an interplanetary constant. 

Luckily, framing technosignatures in terms of environmental modification can quickly pay off if we move from the case of human design towards a broader evolutionary paradigm that deals with environmental modifications caused by life at large (i.e., intelligent or not)---\textit{niche construction theory}. As codified by \cite{laland_introduction_2016}, ``niche construction refers to the modification of selective environments by organisms'', and it offers a theory of environmental modification that is not exclusive to technology. Essentially, niche construction is something that living systems have been doing for billions of years in very detectable ways. It is not just humans, ants, and beavers that bring artificiality into the world---\cite{laland_introduction_2016} mentions snails, worms, corals, seabirds, all of them doing that in order to shape evolutionary pressures in their environment, acting essentially as agents that co-produce their own evolutionary trajectory \citep{walsh_organisms_2015}. Importantly, this view grants organisms in general a kind of agency that is \textit{constitutive} of the niches organisms operate in, meaning that these environments can be reliably recognized as symptoms of organisms that construct them, because the constructed niches are essentially results of their ``life activities'' \citep{lewontin_gene_2001}.

Following this line of reasoning, one may concur that what SETI calls technosignatures---detectable environmental modifications by technological agents---turns out to be a special case of what evolutionary biologists call niche construction---detectable environmental modification by living agents (while remaining agnostic about the exact biochemical makeup of the biosphere they belong to). If we take this seriously, then the distinction between bio- and technosignatures begins to look less like a principled boundary and more like an artifact of disciplinary convention. At the limit, any technosignature is a biosignature (since it is contingent on the existence of extraterrestrial forms of intelligent life), and under a sufficiently generous definition, the reverse may hold too: any case of niche construction could count as a technosignature.

\section{Substrate-agnostic ecology}

We thus arrive at the third level of the substrate-agnostic perspective, one in which biosignatures and technosignatures converge under the umbrella of \textit{ecology}: ``the entire science of the relationships of the organism to its surrounding external world'' \citep[p.~286]{haeckel_generelle_1866}. Niche construction theory, as integrated within the Extended Evolutionary Synthesis \citep{laland_extended_2015}, offers the conceptual apparatus needed to flesh out this intuition about ecology as a science of relations in substrate-agnostic terms.

The key insight is that niche construction operates through \textit{reciprocal causation}: organisms modify their environments, and those modified environments in turn shape the selection pressures acting back on organisms \citep{laland_extended_2015}. This reciprocity implies that neither agent nor environment is ontologically prior. Instead, they are co-constitutive, meaning that inasmuch as the identity of a niche-constructing system is always partially determined by the environment it has constructed, its relevant ``substrate'' is never purely internal but is distributed across the organism-environment coupling. Furthermore, niche construction introduces a channel of inheritance that operates alongside genetic transmission: modified environments persist across generations, creating what \cite{laland_introduction_2016} call \textit{ecological inheritance}---the legacy of ancestral niche-constructing activity that biases the developmental and selective conditions encountered by descendants. This form of inheritance is substrate-agnostic by nature, since what matters is the informational structure of the modified environment, not the biochemical constitution of the agents who produced it. In this context, niche construction also touches upon the phenomenon of \textit{stigmergy}, i.e., communication and coordination through environmental modification \citep{heylighen_stigmergy_2016,grasse_reconstruction_1959}. Interpreted substrate-agnostically, any system in which agents read from and write to a shared medium, thereby modulating the conditions for subsequent agents, exhibits stigmergic dynamics \citep[cf.][]{walsh_organisms_2015}. At planetary or extra-planetary scales, the atmospheric, geological, chemical, or other micro-/macrostructural modifications produced by different bio-/technological agents constitute stigmergic traces, which is precisely what concepts of bio-/technosignatures aim to capture. From this vantage point, the distinction between what is biological and what is technological at best reduces to case-by-case questions about the type of niche-constructing agents responsible for the given modification, rather than about any fundamental ontological difference in the kind of modification itself. At worst, it becomes irrelevant. Alternatively, the kind of context one may need to resolve the ontological status of the given modification may never be available at a sufficiently granular level.

Niche construction---whether through technology-as-we-know-it or through agentic environmental modification in general---lies at the heart of an organism's relationship to its environment, and hence represents an ecological phenomenon in the Haeckelian sense. \cite{walsh_organisms_2015} also defines agency as an ecological phenomenon that lies in a system's capacity to pursue goals by responding to the affordances of its environment. This definition does not presuppose any specific substrate---it only requires agents and environments, plus the relationships between them. Thus, one can generalize on Walsh's definition of agency and simply talk about \textit{substrate-agnostic ecology}, which offers a framework centered on agents, not their nature (i.e., biological, technological, hybrid, or whatsnot). Ecological relationships of agents can thus also be abstracted from the particularities of any single planetary arrangement. 

\section{Conclusion}

Substrate-agnostic ecology proposes that the available conceptualizations of technosignatures and biosignatures share a common theoretical assumption: the existence of niche-constructing agents within planetary environments. The three steps of abstraction outlined in this paper---from substrates of life, substrates of technology, and from the partition between technosignatures and biosignatures---open up a space for thinking about habitability that is no longer tethered to terrestrial conditions. By treating agency as broadly distributed across ecologies, we can begin to think about habitability, planetary evolution, and the search for extraterrestrial intelligence in a genuinely cosmic way. In this manner, substrate-agnostic ecology expands on the insights of the lyfe/genesity framework \citep{wong_searching_2022} by building a unified philosophical perspective. This perspective treats the coevolution of biospheres and technospheres as the central problem for not one, but at least three disciplines: astrobiology, SETI, and environmental sciences broadly conceived (including Earth-system science, ecology, and environmental humanities). While this perspective is only preliminarily sketched in this contribution, it invites future research to examine the position of the concept of intelligence within this framework \citep[especially concerning whether or not it is equivocal with the generic concept of life, cf.][]{aguera_y_arcas_what_2025}.

\section*{Acknowledgments}

\noindent The author used Claude Opus 4.6 (Anthropic) for assistance with compiling the \LaTeX\ version of the manuscript and proofreading the text. All substantive intellectual content, including the arguments, analysis, and conclusions presented in this paper, is solely the author's own.

\end{document}